\documentclass[aps,prl,twocolumn,groupedaddress,showpacs,%
amsmath,preprintnumbers]{revtex4}
\input{amssym}
\usepackage{hyperref} 

\def\t {\tilde}
\def\be {\begin{equation}}
\def\ee  {\end{equation}}
\def\bea {\begin{eqnarray}}
\def\eea {\end{eqnarray}}
\def\nn {\nonumber}
\begin{document}
\preprint{gr-qc/04}
\title{Quantum black holes from null expansion operators}
\author{Viqar Husain and Oliver Winkler}
\email[]{husain@math.unb.ca, oliver@math.unb.ca}
\affiliation{Department of Mathematics and Statistics,
University of New Brunswick, Fredericton, NB, Canada E3B 5A3}
\date{\today}
\begin{abstract}

Using a recently developed quantization of spherically symmetric gravity coupled
to a scalar field, we give a construction of null expansion operators that allow a definition 
of general, fully dynamical quantum black holes. These operators capture the intuitive 
idea that classical black holes are defined by the presence of trapped surfaces, 
that is surfaces from which light cannot escape outward. They thus provide a mechanism 
for classifying quantum states of the system into those that describe quantum black 
holes and those that do not. We find that quantum horizons fluctuate, confirming 
long-held heuristic expectations. We also give explicit examples of quantum black hole 
states. The work sets a framework for addressing the puzzles of black hole physics in a 
fully quantized dynamical setting.
  
\end{abstract}

\maketitle
 
Black holes are probably the most fascinating prediction of general relativity. A host of 
observational data over the last decade has firmly established them as objects within 
experimental reach \cite{rn,mr}. From a theoretical point of view, they are particularly 
attractive in that they offer a glimpse into the as yet unknown realm of quantum gravity. 
Examples are the analogy of black hole mechanics with the laws of thermodynamics,  Hawking 
radiation, and the question of the final fate of the matter whose collapse forms the black 
hole. The latter is closely connected with the question of what happens to the 
classical singularity. 

While classical and semiclassical methods have yielded important insights \cite{rw}, there is no 
doubt that investigations in a fully quantized setting are required to obtain a deeper understanding 
of black hole physics. As a first step in this direction, we developed a quantum theory of 
spherically symmetric gravity systems \cite{hwbh1}. This has the attractive feature that the 
classical singularity is resolved at the quantum level. In order to use this framework for 
studying dynamical black holes, a practically useful definition of quantum black holes is 
required, as well as a criterion for selecting the associated quantum states. This is the topic 
we address in this Letter. 

Traditionally, the notion of a black hole has been intimately connected with the idea of an 
event horizon. The reason is that this is a very useful concept to study static aspects of black 
holes, and has indeed been used to derive important results, some of which were mentioned 
above. Its drawback, however, is its global nature, which makes it unsuitable for investigating 
dynamical processes, such as gravitational collapse, black hole mergers, and black hole evaporation. 
To remedy this situation, more local definitions of a classical black hole have been developed. 
They are based on the concept of a trapped spatial 2-surface.

The intuitive idea is to look at the light emanating from a surface. If no light propagates outward 
(defined in a suitable sense), then the surface is considered to be trapped. This criterion 
may be used to divide regions of space into those containing trapped surfaces, and those 
that do not. The outermost boundary of a trapped region may be used as a definition of the 
(dynamical) black hole horizon.   
   
To make this idea more precise consider a closed embedded 2-surface $S$ with normal $s^a$ 
($s_a s^a=1$) in a spatial slice $\Sigma$ of a spacetime $(M,g_{ab})$. Let $n^a$ ($n^an_a=-1$) 
denote the timelike normal to $\Sigma$ as embedded in $M$. Then we have the decompositions 
\be 
q_{ab} = g_{ab} + n_a n_b, \ \ \ \ \sigma_{ab} = q_{ab} - s_a s_b, 
\ee
for the induced metrics $q_{ab}$ on $\Sigma$ and $\sigma_{ab}$ on $S$. The expansions
of the future pointing "outgoing" $(+)$ and "ingoing" $(-)$ null vectors 
$l_{\pm}^a = n^a \pm s^a$ on $S$ are defined by 
\be 
 \theta_{\pm} = \sigma^{ab}D_a\ l_{\pm b} = s_as_b\pi^{ab} \pm D_as^a 
\label{theta}
\ee  
where $D_a$ is the $q_{ab}$ covariant derivative, and the last equality uses the 
Arnowitt-Deser-Misner ADM momentum variable $\pi^{ab}=K^{ab}-Kq^{ab}$, where $K^{ab}$ is the 
extrinsic curvature of $\Sigma$ and $K$ is its trace. The surface $S$ is called {\it outer trapped} 
if 
\be 
\theta_+ \le 0,\ \ \ \ \theta_-< 0,
\ee 
and {\it marginally outer trapped} if 
\be 
\theta_+ = 0, \ \ \ \ \theta_-< 0.  
\label{margot}
\ee
These conditions hold for generic slices of all known black hole spacetimes, and therefore provide 
a natural test to see if ($\Sigma, q_{ab}, \t{\pi}^{ab}$) is a slice of a black hole spacetime. The time evolution of the marginal outer trapping boundary forms a 3-surface in $M$ called the {\it trapping
horizon} \cite{sh}. Unlike the event horizon, it is not necessarily a null surface. This reflects its usefulness for capturing dynamical processes which change the size and location of a horizon.
The intersections of a trapping horizon with the spatial hypersurfaces of a canonical decomposition 
provide a very useful tracking mechanism for such processes in time sequence. This is routinely used in numerical evolution for example. It is also the appropriate framework for our canonical quantization 
program.

The system we consider is a canonical formulation of the asymptotically flat gravity in spherical symmetry. 
Although this system has been studied both with and without matter classically and semi-classically 
\cite{bcmn,hawk,unruh,hajicek,tk,kuchar,louko,grum}, a complete quantization with scalar field coupling has to 
our knowledge never been achieved. Recently we have given a new  quantization of this model in Painleve-Gullstrand coordinates 
\cite{hwbh1}. Our main goal in the present work  is to provide in that framework a construction 
of operators $\hat{\theta}_\pm$ corresponding to the classical expansions $\theta_\pm$, and use them to 
identify black hole quantum states. 

We first review briefly the canonical setting and its quantization, and then proceed to describe in 
detail a construction of  expansion operators. Using these operators, we give a criterion for defining 
the quantum states that correspond to classically trapped configurations, and construct some explicit 
examples of such states.  

The classical Hamiltonian theory of the spherically symmetric gravity-scalar field model used is 
developed in \cite{hwbh1}. This system is adapted to the Painleve-Gullstrand (PG) coordinates, in which
the black hole metric is 
\be 
ds^2 = -dt^2 + \left(dr + \sqrt{\frac{2M}{r}} dt \right)^2 + r^2 d\Omega^2.
\label{PG}
\ee
The spatial metric $e_{ab}$ given by the constant $t$ slices is flat and the extrinsic curvature 
of the slices is 
\be 
K_{ab} = -\sqrt{\frac{2M}{r^3}}\left( e_{ab} - {\frac{3}{2}} s_a s_b\right), 
\label{K}
\ee
where $s^a=x^a/r$ for Cartesian coordinates $x^a$. It is evident from this that the black hole 
mass information is contained only in the extrinsic curvature, which in the  ADM
phase space variables $(q_{ab},\tilde{\pi}^{ab})$ determines the momenta $\tilde{\pi}^{ab}$ 
conjugate to the spatial metric $q_{ab}$. 

The phase space of the model is defined by prescribing a form of the gravitational 
phase space variables $q_{ab}$ and $\tilde{\pi}^{ab}$.  Asymptotic flatness falloff conditions 
are then imposed on these variables, and on the lapse and shift functions $N$ and $N^a$, such 
that the canonical ADM action is well defined.  

In this setting, the following parametrization for the 3-metric and conjugate momentum gives a 
reduction to spherical symmetry \cite{hwbh1}
\bea
q_{ab} &=& \Lambda(r,t)^2\ s_a s_b + \frac{R(r,t)^2}{r^2}\ ( e_{ab} - s_a s_b)\\
\t{\pi}^{ab} &=& \frac{P_\Lambda(r,t)}{2\Lambda(r,t)}\ s^a s^b + \frac{r^2 P_R(r,t)}{4R(r,t)}\ 
(e^{ab} - s^a s^b).
\label{reduc}
\eea
Substituting these into the 3+1 ADM action shows that the pairs $(R,P_R)$ and $(\Lambda,P_\Lambda)$ 
are canonically conjugate variables.  The reduced ADM 1+1 field theory action takes the form   
\bea 
S_R &=& \frac{1}{2G}\int dtdr \left(P_R\dot{R} + P_\Lambda\dot{\Lambda} + P_\phi\dot{\phi}  
    \right)\nn\\
     && +\ {\rm constraint\ and\ surface\ terms}.
\eea
This completes the definition of the classical theory. The two reduced constraints, the Hamiltonian 
and radial diffeomorphism generators are first class. A surface term is needed to ensure 
functional differentiability of $S_R$.   

The classical expansions (\ref{theta}) in spherical symmetry are functions of the 
phase space variables given by  
\be
\theta_\pm = -\frac{1}{2\Lambda}\left(2R^2\Lambda \Lambda' \pm P_\Lambda + 4\Lambda^2 RR'\right).  
\label{hor}
\ee
Given data $(R(r), P_R(r);\Lambda(r),P_\Lambda(r))$ on a spatial hypersurface $\Sigma$, the marginal 
trapping horizon(s) are located by finding the solution coordinates $r=r_i$ ($i=1\cdots n$) 
of Eqns. (\ref{margot}) (since in general there may be more than one solution). 
The corresponding radii $R_i= R(r_i)$ are then computed. The size of the horizon on the slice 
$\Sigma$ is the largest value in the set $\{R_i\}$.  

Turning now to the quantum theory where the horizon operators corresponding to these classical 
conditions will be defined, let us recall the quantization used in \cite{hwbh1}.
A typical basis state is given by 
\bea 
&&|e^{i \sum_k a_k P_R(x_k)}, e^{i\sum_l b_l P_\Lambda(y_l)/L},e^{i L^2\sum_m c_m P_\phi(z_m)}\rangle \nn\\
&&\equiv
|a_1\ldots a_{N_1}; b_1\ldots b_{N_2},c_1\ldots c_{N_3}\rangle, 
\eea 
where $a_k \cdots c_m$ are real numbers which are the analogs of the occupation numbers in 
a Fock space representation, and the factors of $L$ in the exponents reflect the length 
dimensions of the respective field variables. The inner product on this basis is given by 
\be
\langle a_1 \ldots c_{N_3}| a'_1\ldots  c_{N_3}'\rangle  
= \delta_{a_1,a_1'}\cdots \delta_{c_{N_3},c_{N_3}'}, 
\ee
if the states contain the same sampled points, and is zero otherwise. 

The configuration operators are defined by
\be 
\hat{R}_f\ |a_1  \ldots c_{N_3}\rangle = L \sum_k a_k f(x_k)|a_1  \ldots c_{N_3}\rangle,
\label{Rf}
\ee
which is the operator corresponding to the classical smeared variable 
$R_f= (1/L)\int_0^\infty f Rdr$, for suitable test functions $f$ -- (the factor $L$ gives 
it dimension length). The conjugate momentum operator exists only in the exponentiated form 
in this quantization. Its action is   
\bea
&& \widehat{e^{i \lambda_j P_R(x_j)}}|a_1\ldots   c_{N_3}\rangle
  \nn\\
&&=|a_1\ldots, a_j-\lambda_j,\ldots   c_{N_3}\rangle
\label{trans}
\eea
where $a_j$ is $0$ if the point $x_j$ is not part of the original basis state. In this 
case the action creates a new `excitation' at the point $x_j$ with `mode'
$-\lambda_j$. These definitions give a faithful representation of the classical Poisson 
bracket of these variables. The Poisson bracket commutator correspondence fixes   
$L = \sqrt{2}\ l_P$, where $l_P$ is the Planck length. Similar definitions and the 
commutator result hold for the other phase space variables $\Lambda_f$, exp$(iP_\Lambda/L)$
and $\phi_f$, exp$(iL^2P_\phi)$, where the corresponding operators act on the basis state 
excitations $b_1\ldots b_{N_2}$ and $c_1\cdots c_{N_3}$ respectively. In the following 
we suppress the scalar field part of the basis states, since these are not needed for 
defining operator analogs of $\theta_\pm$.  

With the basic operators defined, we are in a position to address the question of the horizon 
operators. This requires defining operators for $P_\Lambda(r,t)$, $R'(r,t)$, and $\Lambda'$ 
as a glance at Eqns. (\ref{hor}) shows. Since only translation operators (\ref{trans}) 
are available in our quantization, and these are not weakly continuous in the parameter 
translation parameter $\lambda$ \cite{hwbh1}, we define $P_\Lambda$ indirectly by 
\be 
  \hat{P}_\Lambda^\lambda = \frac{l_P}{2i\lambda}\ \left( \hat{U}_\lambda - \hat{U}_\lambda^\dagger\right)
\ee
where $0<\lambda\ll 1$ is an arbitrary but fixed parameter, and $U_\lambda$ denotes 
exp$(i\lambda P_\Lambda/L)$. This is motivated by the corresponding classical expression, 
where the limit $\lambda\rightarrow 0$ exists, and gives the classical function $P_\Lambda$. 
$\lambda$ is perhaps best understood as a ratio of two scales, $\lambda = l_p/l_0$, 
where  $l_0$ is a system size. It is evident that momentum in this quantisation can be given 
approximate meaning only for $\lambda\ll 1$. $\lambda$ is also the minimum value 
by which an excitation can be changed. 
 
Definitions for the operators corresponding to $R'$ and $\Lambda'$ are obtained by implementing 
the idea of finite differencing using the operator $\hat{R}_f$ (\ref{Rf}). We use narrow 
Gaussians with variance proportional to the Planck scale, peaked at coordinate points 
$r_k+\epsilon l_P$, where $0<\epsilon \ll 1$ is a parameter designed to sample neighbouring points: 
\be 
f_{\epsilon}(r,r_k)= \frac{1}{\sqrt{2\pi}}\ {\rm exp} \left[-\ \frac{(r-r_k-\epsilon l_P)^2}
{2 l_P^2}\right]   
\ee
Denoting $R_{f_\epsilon}$ by $R_{\epsilon}$ for this class of test functions we define  
\be 
  \hat{R'}(r_k):= \frac{1}{l_P \epsilon}\ \left( \hat{R}_{\epsilon} - \hat{R}_{0} \right).
\label{R'}
\ee
A further motivation of this form is that in the gauge $R=r$ the corresponding classical expression 
gives unity in the limit $\epsilon \rightarrow 0$. This definition captures the simplest finite difference approximation to the derivative at the operator level. If a state has excitations separated by $\epsilon l_P$ then the classical intuition of the derivative is approximately implemented for that state. Such states may be viewed as potential "pre-semiclassical" states in the sense that they represent approximately differentiable field values, as opposed to "purely quantum" states which may be "nowhere differentiable" in the sense that they represent isolated excitations at large coordinate separations. The quantisation of 
$\Lambda'$ is analogous to (\ref{R'}). 

Apriori the parameters $\lambda$ and $\epsilon$ have no relation to each other since 
the former is an element of radial field excitation, and the latter is a coordinate separation. 
However, the classical falloff $R(r)\sim r$ for large $r$ suggests that it is plausible to set 
$\lambda=\epsilon$. Nevertheless we will keep them separate for the present discussion. 

Lastly, we note that the classical marginal trapping conditions (\ref{margot}) are independent 
of the overall factor of $1/\Lambda$ in (\ref{hor}), for $\Lambda >0$. A classically equivalent 
expression of these conditions is obtained by multiplying (\ref{hor}) by $1=\Lambda/\Lambda$, 
taking the numerator inside the bracket, and dropping the factor $1/\Lambda^2$. This gives the 
classical form we utilise for the quantum theory.   

Putting all these pieces together, we can construct the local operators 
\bea
\hat{\theta}_\pm(r_k) &=& -\frac{2}{\epsilon l_P} \hat{R}_0^2 \hat{\Lambda}_0^2 
\left(\hat{\Lambda}_\epsilon - \hat{\Lambda}_0 \right) 
\mp  \frac{l_P}{2i\lambda}\left( \hat{U}_\lambda - \hat{U}_\lambda^\dagger \right)\hat{\Lambda}_0 \nn\\
&&- \frac{4}{\epsilon l_P}\hat{\Lambda}_0^3 \hat{R}_0 \left(\hat{R}_\epsilon - \hat{R}_0 \right), 
\eea
which have a well defined action on the basis states. We now use these operators 
to identify quantum black hole states. 

A first observation is that our basis states are not eigenstates of $\hat{\theta}_\pm$, since the 
$\hat{U}_\lambda$ are finite shift operators. One might be tempted to look for solutions of the condition 
$\hat{\theta}_+|\Psi> =0$, which looks like a Dirac quantization condition. However, as in the latter case, 
the resulting solutions may not be normalizable states. For this reason we propose that {\it a state 
$|\Psi\rangle$ represents a quantum black hole} if
\be
\langle \Psi| \hat{\theta}_+(r_k) |\Psi\rangle = 0, \ \ \ \langle \Psi| \hat{\theta}_-(r_k) |\Psi\rangle <0. 
\label{qbh}
\ee
for some $r_k$. The corresponding {\it horizon size} is given by 
\be 
R_H = \langle \Psi| \hat{R}_{0} | \Psi\rangle. 
\ee 
This proposal gives a simple two step  procedure for locating a quantum horizon: Given a state, 
one scans all its excitations using $\hat{\theta}_+(r_k)$ until a location is found where the 
conditions (\ref{qbh}) hold. Once this is done, the physical horizon size is determined by computing 
the expectation value of $\hat{R}_0(r_k)$. This resembles the 
classical horizon location procedure.  

It is apparent that the coordinate location of the quantum horizon $r_k$ is sharp in the sense that 
it is determined by the peak of a Gaussian. However, the horizon size is not sharp since the states 
satisfying the horizon conditions are not eigenstates of $\hat{R}_f$. As a consequence the horizon 
area is also not sharp. This has clear implications for the semiclassical entropy--area equation, 
which in light of this quantization will likely arise as an equality of expectation values. 

Let us consider now the effect of the radial diffeomorphisms on the basis states. Denoting their 
action on coordinates by $\phi: r \longrightarrow \phi(r)$, the induced action on a basis element   
$|\mu_1(r_1)\cdots\mu_N(r_N)\rangle$ is 
\be
\hat{\phi}|\mu_1(r_1)\cdots \mu_N(r_N)\rangle)=
|\mu_1(\phi(r_1))\cdots \mu_N(\phi(r_N))\rangle.\nn
\ee
From this it is clear that if a general state $|\Psi \rangle$ satisfies the horizon conditions with 
respect to $\hat{\theta}_\pm(r)$ then $\hat{\phi}|\Psi \rangle$ will satisfy the condition for 
$\hat{\theta}_\pm(\phi(r))$. It follows that the horizon coordinate location in the background 
manifold changes, while the horizon size remains the same. This mirrors exactly what happens at 
the classical level.

Explicit expressions for states satisfying the quantum black hole conditions are not hard to construct.
A first example is the  "vacuum" $|00\cdots00\rangle$ which has no excitations of any of the fields.
It is easy to verify that for this state $\langle \hat{\theta}_\pm(r_k) \rangle = 0$ for all 
coordinates $r_k$, and that  $\langle \hat{R}_0(r_k) \rangle = 0$. The trapping evident in this state 
is trivial in the sense that the expectation values of the terms in $\hat{\theta}_\pm$ are each 
separately zero. 
 
To construct a class of non-trivial black hole states, we note that $\hat{\theta}_\pm$ depends on 
the fixed parameters $\lambda,\epsilon$, and acts on field excitations at the coordinate values 
$r_k$ and $r_k+\epsilon l_p$. Specifically let  
\bea 
|\Psi\rangle &=& \frac{1}{\sqrt{2}}\left( | \cdots a_k,a_{k+1},\cdots; \cdots ,b_k,b_{k+1}, \cdots\rangle \right.\nn\\
&& \left.+ i|\cdots a_k,a_{k+1}\cdots;\cdots b_k-\lambda,b_{k+1},\cdots \rangle \right) 
\label{2state}
\eea 
for excitations $a_k,a_{k+1}$ etc. at locations $r_k$ and $r_{k+1}$. For this state we have 
$\langle \hat{R}_{0}(r_k) \rangle = \sqrt{2}l_P a_k$, and  
\bea
\frac{1}{l_P}\ \langle \hat{\theta}_\pm(r_k) \rangle 
&=& -\frac{8a_k^2}{\epsilon}\left[b_k^2\delta b_k + (b_k-\lambda)^2(\delta b_k+\lambda)\right]\nn\\
&& - \frac{8a_k\delta a_k}{\epsilon} \left[b_k^3 + (b_k-\lambda)^3 \right]\nn\\  
&& \pm\left(\frac{b_k}{2\lambda}-\frac{1}{4} \right),
\label{<theta>}
\eea
where $\delta a_k = a_{k+1}-a_k$, with the same for $b_k$. It is evident from this equation that 
states satisfying the quantum black hole conditions depend on the relative values of the 
excitations at neigbouring points. For illustration let us consider some specific cases.  
(i) $a_k=b_k=0$: $\langle \theta_+(r_k)\rangle >0$ and  
$\langle \theta_-(r_k)\rangle <0$, so there is no horizon. (ii) $b_k =0$: there
is a value of $a_{k+1}(a_k,\lambda,\epsilon)$, given by a linear equation, for which 
$\langle \theta_+(r_k)\rangle = 0$ and  $\langle \theta_-(r_k)\rangle <0$. (iii) More 
generally, the quantum trapping condition $\langle \theta_+(r_k)\rangle = 0$ may be viewed 
as a linear equation either for $a_{k+1}$, or for $b_{k+1}$, with all other corresponding 
values fixed. (iv) The generic state of this type represents normal untrapped 
space with no horizons because the third term in brackets in (\ref{<theta>}) is positive 
and large for $b_k$ of order unity since $\lambda \ll 1$.  The $\hat{\theta}_\pm$ operators will 
have the same practical utility for semiclassical states, which arise as infinite linear 
combinations of the basis states, with a parameter specifying the classical mass.   

Finally it is interesting to note that changing the state (\ref{2state}) by moving the 
$i$ factor from the second component to the first results in an overall sign change in the 
third term in (\ref{<theta>}), with no effect on the first two terms. This means that 
one can obtain white hole states satisfying $\langle \hat{\theta}_+(r_k) \rangle = 0$  
and $\langle \hat{\theta}_-(r_k) \rangle >0$. This shows the richness evident in the state 
space already at the two component level.  

To summarise, let us recall the most prominent results of this work. We have constructed    
operator analogs of the classical null expansions, and used these to define the notion of a 
quantum black hole. The resulting quantum states are normalizable and thus bona fide 
elements of the Hilbert space. Our use of trapping conditions to define  quantum black hole 
states corresponds to the classical dynamical trapping horizon introduced in \cite{sh}, 
which applies to fully dynamical horizons. This is unlike the use of the isolated horizons 
\cite{mb} which are not applicable to dynamical situations. Furthermore, our quantization is 
in a setting that has been shown to be singularity free \cite{hwbh1}. An important and 
physically appealing consequence of our results arises from the fact that the horizon operator 
does not commute with the area operator:  the horizon size is subject to quantum fluctuations 
in the sense that it does not have sharply defined area eigenvalues.

Finally, it may prove valuable that our setup is close to the one used in numerical black hole 
calculations: the Hamiltonian evolution of a given initial state with no quantum horizon can be 
followed to  horizon formation, and beyond. This analogy will facilitate numerical implementation 
of more involved calculations. The stage is thus set to study various dynamical processes of quantum 
black holes such as formation, mergers and evaporation. The first of these processes would be 
the quantum analog of the classical black hole formation in this system studied in \cite{chop}.

\acknowledgments{This work was supported in part by the Natural Science and Engineering 
Research Council of Canada. We thank Arundhati Dasgupta for comments on the manuscript.}

\end{document}